\documentclass[runningheads,a4paper]{llncs}
\usepackage{amssymb}
\usepackage{amsmath}
\usepackage[dvipdfmx]{graphicx}
\usepackage{url}
\newcommand{\keywords}[1]{\par\addvspace\baselineskip
\noindent\keywordname\enspace\ignorespaces#1}

\begin{document}
\mainmatter  
\title{
Detection of non-self-correcting nature of information cascade
}
\titlerunning{}
\author{
     Shintaro Mori \inst{1}\footnote{corresponding author, mori@sci.kitasato-u.ac.jp}
\and Masafumi Hino \inst{2} 
\and Masato Hisakado \inst{3}
\and Taiki Takahashi \inst{4,5}
}
\authorrunning{S. Mori et al.}
\institute{Department of Physics, Kitasato University\\
1-15-1, Kitasato, Sagamihara, Kanagawa 252-0373, Japan\\
\and
NEC  Corporation\\
Siba 5-7-1, Minato-ku, Tokyo 108-8001, Japan
\and
Financial Services Agency\\
Kasumigaseki 3-2-1, Chiyoda-ku, Tokyo 100-8967, Japan\\
\and
Department of Behavioral Science, Faculty of Letters
\\
\and
Center for Experimental Research in Social Sciences, Hokkaido University
\\
Kita 10, Nishi 7, Kita-ku, Sapporo, Hokkaido 060-0810, Japan
}

%
%

\toctitle{Lecture Notes in Computer Science}
\tocauthor{Authors' Instructions}
\maketitle

\begin{abstract}
We propose a method of detecting  
non-self-correcting information cascades 
in experiments in which
 subjects choose an option sequentially by observing the choices 
of previous subjects.
The method uses the correlation function $C(t)$ between the first and 
the $t+1$-th subject's choices. $C(t)$ measures the strength of the 
domino effect, and the limit value $c\equiv \lim_{t\to \infty}C(t)$ 
determines whether the domino effect lasts forever $(c>0)$ or not $(c=0)$. 
The condition $c>0$ is an adequate condition for a non-self-correcting 
system, and 
the probability that the majority's choice remains wrong 
in the limit $t\to \infty$ is positive. 
We apply the method to data from two 
experiments in which $T$ subjects answered two-choice questions:
(i) general knowledge questions ($T_{avg}=60$) and 
(ii) urn-choice questions ($T=63$). 
 We find $c>0$ for difficult questions in (i) and all cases in (ii), 
 and the systems are not self-correcting.
\end{abstract}

\section{Introduction}
 Herding phenomena are ubiquitous in human and animal 
 behavior \cite{Cas:2009,Gra:2014}.   
 An example is an information cascade, in which 
 a person observes others' choices and 
 chooses the majority's choice even though the person's private signal 
 contradicts it \cite{Bik:1992,Dev:1996}. 
 It is a rational behavior 
 for people who are uncertain about choosing. If an information
 cascade occurs, the same mechanism applies to later decision-makers, and 
 the majority's choice tends to prevail. In some cases, the successive 
 choices are wrong, and the cascade leads to irrational herding 
 behavior \cite{Lee:1993}.  
 
 An experimental setup demonstrates a situation in which an information cascade 
 occurs \cite{And:1997}. 
 There are two urns, A and B, and urn A (B) contains two {\it a} ({\it b}) balls
 and one {\it b} ({\it a}) ball. In each run of the experiment, 
 an urn is randomly chosen initially and called X. 
 Then, the subjects guess whether urn X 
 is A or B and choose sequentially. 
 They get a reward for the correct choice.
 In the course of the experiment, each subject draws a ball from X, which 
 is his private 
 signal. If the ball is {\it a} ({\it b}), urn X is 
 more likely to be A (B). 
 He also observes the choices of the previous subjects.
 If the difference between the numbers of subjects who choose each urn
 exceeds two, the private signal cannot overcome the majority's choice. 
 An information cascade starts if someone chooses the majority's 
 choice although his private signal suggests the minority's one.
 As the probability that the first two persons both choose the wrong 
 option is non-zero, the probability for the onset of a cascade  
 where the majority's choice is wrong is positive. 

  We now consider whether the wrong cascade continues \cite{Lee:1993}.
  If it continues forever, the majority's choice converges to 
  the wrong option.
  Information cascades were initially considered to be fragile phenomena. 
  As the trigger of the  
  cascade is a small imbalance, people can be dissuaded from following 
  the majority's choice \cite{Bik:1992}. 
  In addition, an agent model 
  with a Bayesian update of the private belief showed that the 
  information cascade is self-correcting \cite{Goe:2007}.
  As the number of agents tends toward infinity, the wrong cascade 
  disappears, and the majority's choice converges to the optimal option.   

  Using an information cascade experiment with a general knowledge 
  two-choice quiz, we have shown that a phase transition occurs between 
  a one-peak phase 
  and a two-peak phase \cite{Mor:2012}. If the questions are easy, 
  the ratio $z(t)$ of the correct choices of 
  $t$ subjects converges to a value $z_{+}>1/2$ in the limit $t\to \infty$. 
  As there is only one peak in the probability distribution 
  function of $z(t)$, 
  we call the corresponding phase the one-peak phase \cite{His:2011,His:2012}. 
  If the questions are difficult and most people do not know 
  the answers, $z(t)$ converges to $z_{+}>1/2$ or $z_{-}<1/2$. 
  One cannot predict the value in 
  $\{z_{+},z_{-}\}$ to which $z(t)$ converges. We call the 
  corresponding phase the two-peak phase.
  In the two-peak phase, the wrong cascade does not necessarily 
  disappear, and the system is not self-correcting.

  It was recently shown that the limit value of the normalized 
  correlation function 
  is the order parameter of the phase transition \cite{Mor:2015}. 
  The normalized correlation function 
  shows how the first subject's choice propagates to later subjects. 
  It provides a 
  measure of the domino effect. In addition, the positiveness of the 
  limit value 
  is a sufficient condition for a non-self-correcting system. 
  By extrapolating the results for a finite system to infinity, 
  we can determine whether the system is self-correcting. 
  We report on the application 
  of the method to data from two types of information cascade experiments. 
  In section 2, we define the normalized correlation function.
  We also explain the behavior of the function in each phase and 
  the extrapolation method 
  used to estimate its limit. 
  We present the results of the data analysis in section 3. 
  Section 4 summarizes the results. 

\section{Correlation function and asymptotic behaviors}
 We consider a typical information cascade experiment.
 $T$ subjects answer a two-choice question sequentially in each run.
 We denote the order of the subjects as $t$, where $t=1,2,\cdots,T$.
 We denote the choice of subject $t$ 
 by $X(t) \in \{0,1\},t=1,2,\cdots,T$.
 If the choice is true (false), $X(t)$ takes 1 (0).
  
\begin{figure}
\centering
\includegraphics[width=12cm]{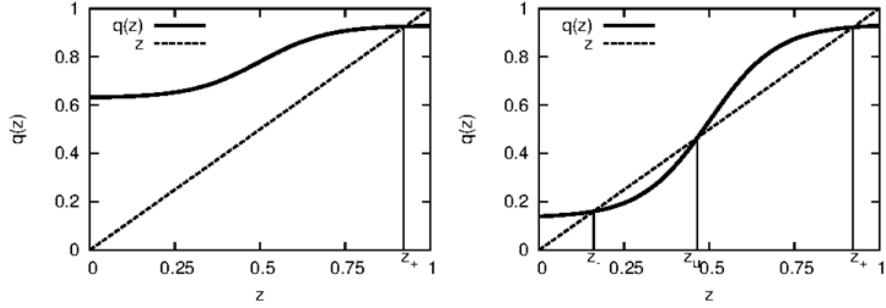}
\caption{Response function $q(z)$ vs. $z$.
 Left panel shows the one-peak phase, in which
 there is one solution, $z_{+}$, for $z=q(z)$.
Right panel shows the two-peak phase, in which
there are three solutions, $z_{-}<z_{u}<z_{+}$, for
$z=q(z)$.
}
\label{fig:ONE-TWO}
\end{figure}

The correlation function $C(t)$ is defined 
as the covariance between $X(1)$ and $X(t+1)$ divided by the variance of $X(1)$:
\[
C(t)\equiv \mbox{Cov}(X(1),X(t+1))/\mbox{Var}(X(1)).
\]
$C(t)$ can be expressed as the  difference 
of two conditional probabilities.
\begin{equation}
C(t)=\mbox{Pr}(X(t+1)=1|X(1)=1)-\mbox{Pr}(X(t+1)=1|X(1)=0) \label{eq:rC}.
\end{equation}
$C(t)$ shows the degree to which the first subject's choice
 is transmitted to later subjects. It is a measure of 
 the domino effect in an information cascade.

 $C(t)$ is generally positive, and its
 asymptotic behavior depends on the phase of the system and 
 the shape of the 
 response function $q(z)$. Here $q(z)$ represents the dependence of 
 the probability of the correct 
 choice by subject $t+1$
 on the ratio $z(t)$ 
 of the correct choices of the previous $t$ subjects. 
\[
q(z)\equiv \mbox{Pr}(X(t+1)=1|z(t)=z)\,\,\, , \,\,\, z(t)=\frac{1}{t}\sum_{s=1}^{t}X(s).
\]
 With the definition of $q(z)$, the stochastic process $\{X(t)\},t=1,2\cdots$ 
 becomes a generalized P\'{o}lya urn process \cite{Hil:1980}.
 If there is one solution for $z=q(z)$ at $z_{+}$ (left panel 
in Fig.\ref{fig:ONE-TWO}), $z(t)$ converges to $z_{+}$.
$C(t)$ shows power-law decay for large $t$ with two constants, $c'$ and $l$, as
\[
C(t) \simeq c'\cdot t^{l-1}\,\,\,\ l<1.
\]
 Here, $l$ is the exponent for the power-law decay and is less than 1.
 The value of $l$ is given by $g'(z_{+})$ \cite{Hod:2004,His:2012}.
If there are three solutions for $z=q(z)$ at $z_{-}<z_{u}<z_{+}$ (right 
panel in Fig.\ref{fig:ONE-TWO}), the system is in the two-peak phase;
$\lim_{t\to \infty}z(t)=z_{+}\,\,\, \mbox{or}\,\,\,z_{-}$ \cite{Hil:1980}. The limit value 
$c\equiv \lim_{t\to\infty}C(t)$ is positive, and 
 the first subject's 
 choice propagates to an infinite number of later subjects \cite{Mor:2015}. 
$C(t)$ behaves asymptotically as
\begin{equation}
C(t) \sim c+c'\cdot t^{l-1} \label{eq:ct}.
\end{equation}
Here $c'\cdot t^{l-1}$ is the subleading term of $C(t)$, and $l$ is 
 given by the larger value among $\{g'(z_{+}),g'(z{-})\}$.
 Further, $c$ acts as an order parameter of the phase transition, and
 eq.(\ref{eq:ct}) is the general asymptotic behavior of $C(t)$ \cite{Mor:2015a}.

As it is difficult to estimate $c$ using $c\equiv \lim_{t\to\infty}C(t)$
 with empirical data, where the system size and number of samples are strictly limited, 
we introduce two quantities for the estimation. First, we define 
the $n-$th moment $m_{n}(t)$ for 
$C(t)$ as $m_{n}(t)\equiv \sum_{s=0}^{t-1}C(s)(s/t)^{n}$. We define the integrated 
correlation time $\tau(t)$ as $\tau(t)=m_{0}(t)$. We also define the 
second moment correlation time $\xi(t)$ as 
$\xi(t)\equiv t\cdot \sqrt{m_{2}(t)/m_{0}(t)}$.
Using the asymptotic behavior of $C(t)$, 
we estimate the subsequent asymptotic behavior 
of $\tau(t)/t$ and $\xi(t)/t$.
\begin{eqnarray}
\tau(t)/t &\simeq& c+\frac{c'}{l}\cdot t^{l-1} \label{eq:taut} 
\\
\xi(t)/t &\to & 
\begin{cases}
\sqrt{l/l+2}   \,\,\,\, c=0\\
\sqrt{1/3} \,\,\,\, c>0   \label{eq:xit}
\end{cases}
\end{eqnarray}
As $\tau(t)/t$ is defined as the summation 
of $C(s)$ over $0\le s <t$ divided by $t$, 
the standard error becomes smaller than that of $C(t)$.
The asymptotic behavior of $\tau(t)/t$ in eq.(\ref{eq:taut}) provides a more reliable 
estimate of $c$ and $l$ than the fitting of $C(t)$ to eq.(\ref{eq:ct}). 
$\xi(t)/t$ also provides a reliable estimate for $l$ \cite{Mor:2015a}.
If $c>0$, the leading term of $C(t)$ is the constant $c$, and $l$ 
should be interpreted as $l=1$. 
 
 We define whether the system is self-correcting according to 
 whether $z(t)$ always converges to $z_{+}$. In the one-peak (two-peak) 
 phase, the system is (non-)self-correcting.   
 If $c>0$, the system is in the two-peak phase and is non-self-correcting.
 However, $c=0$ does not necessarily mean that the system is self-correcting.
 For the system to be self-correcting, $q(z)=z$ has to have only one 
 solution, $z_{+}$.  

\section{Domino effect and detection of non-self-correcting nature}
 We study the domino effect and non-self-correction 
 in information cascades.
 We discuss two types of information cascade experiments.

 In experiment 1 (EXP-I), subjects answered a general knowledge two-choice quiz.
 First, the subjects answered using only their own knowledge.
 Then, they observed the choices of previous subjects 
 and answered the question again.
 The average length of the sequence of subjects is $T=60$, and 
 the number of choice 
 sequences is $240$. The choice sequences are classified
 into four bins according to the ratio of correct choices $z_{0}(T)$ of the 
 first answers without 
 observation as
 $z_{0}(T)=50\% \pm 5\%,60\% \pm 5\%,70\% \pm 5\%$, and 
 $80\% \pm 5\%$, and the number of samples in each bin is 
 $38(50\% \pm 5 \%),52(60\% \pm 5\%),38(70\% \pm 5 \%),$ and $38(80\% \pm 5 \%)$,
 respectively \cite{Mor:2013}. 

 Experiment 2 (EXP-II) is similar to the situation explained 
 in the Introduction. 
 There are two urns, A and B, which contain {\it a} and {\it b} balls in different 
 configurations. We use two configuration patterns: 
(i) two {\it a} balls and one {\it b} ball in urn A vs. one {\it a} ball and  two {\it b} balls 
 in urn B and (ii) five {\it a} balls and four {\it b} balls in urn A vs. 
 four {\it a} balls and five {\it b} 
 balls in urn B. Urn $X\in \{\mbox{A,B}\}$ is chosen at random  at the beginning of each 
 run, and subjects are asked to choose between A or B.
 Each subject draws one ball from $X$ and checks whether it is {\it a} or {\it b}.
 The ball corresponds to the type of urn $X$ with probability $q=2/3 (5/9)$
 for (i) [(ii)].
 In addition, the subject also observes the choices of 
 previous subjects. Our results, unlike those of previous experiments \cite{And:1997,Kub:2004,Goe:2007}, 
 show the summary statistics of the number of subjects 
 who have chosen each urn.
 The length $T$ and number of questions $I$ are 
 63 and 200, respectively, for $q\in \{2/3,5/9\}$ \cite{Hin:2015}.

 We denote the choice sequences in  each bin as $\{X(i,t)\},i=1,\cdots,I,t=1,\cdots,T(i)$.
 Here, the length of the sequence depends on question $i$ in EXP-I; we denote it as $T(i)$.
 The number of samples $I$ also depends on the bins.
 In EXP-II, $T(i)=63$, and $I=200$.
 First, we estimate $C(t)$ and its standard 
 error $\Delta C(t)$ using eq.(\ref{eq:rC}).
 We denote the estimate and standard error of the probabilities as 
 $q_{x}(t+1)=\mbox{Pr}(X(t+1)=1|X(1)=x)$ and $\Delta q_{x}(t+1)$, respectively.
 They are estimated 
from experimental data $\{X(i,t)\}$ as
\begin{eqnarray}
q_{x}(t+1)&=&
\frac{1+\sum_{i=1}^{I}X(i,t+1)
\delta_{X(i,1),x}}{N_{x}+2}, \nonumber \\
N_{x}&=&\sum_{i=1}^{I}\delta_{X(i,1),x},  \nonumber \\
\Delta q_{x}(t+1)&=&
\sqrt{\frac{q(x,t+1)(1-q_{x}(t+1))}{N_{x}+3}}.
\nonumber 
\end{eqnarray}
 Here, we use the expectation value and standard deviation 
 obtained from the posterior probability distribution for the probabilities. 
$C(t)$ is then estimated as 
\[
C(t)=q_{1}(t+1)-q_{0}(t+1).
\]
 The error bars of $C(t)$ are given as 
\begin{equation}
\Delta C(t)=\sqrt{\Delta q_{1}(t+1)^{2}+\Delta q_{0}(t+1)^{2}}. \label{ct_error}
\end{equation}

 Using  $C(t)$ and $\Delta C(t)$, we estimate the 
error bars of $m_{n}(t)$ as
\[
\Delta m_{n}(t)=\sqrt{\sum_{s=1}^{t-1}\Delta C(s)^{2}(s/t)^{2n}}.
\]
Here we assume that $\Delta C(s)$ and $\Delta C(s')$ are 
independent of each other if $s\neq s'$. 
We estimate the error bars 
of $\tau_{t}(t)$ and $\xi_{t}(t)$ as
\begin{eqnarray}
\Delta \tau_{t}&=&\frac{1}{t}\Delta m_{0}(t),
 \nonumber \\
\Delta \xi_{t}&=& \sqrt{\xi_{t}}(\Delta m_{2}(t)/2m_{2}(t)
+\Delta m_{0}(t)/2m_{0}(t)) .
\end{eqnarray}
In the estimation of $\Delta \xi_{t}$, we assume that 
$\Delta m_{2}(t)$ and $\Delta m_{0}$ are completely correlated.

\subsection{EXP-I: General knowledge quiz case}

\begin{figure}[t]
\centering
\includegraphics[width=9cm]{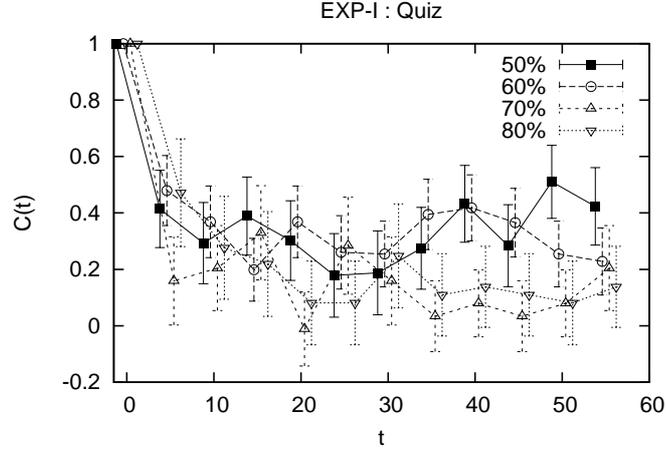}
\caption{$C(t)$ vs. $t$ for EXP-I. The sample choice sequences are
classified according to the value of $z_{0}(T)$ as $z_{0}(T)=50\% \pm 5\% (\blacksquare),60\% \pm 5\% (\bigcirc),
70\% \pm 5\% (\triangle)$, 
and $80\% \pm 5\% (\bigtriangledown)$. 
We plot only data with the interval $\Delta t=5$. To see the behavior clearly, 
 we slightly shift the data horizontally.
}
\label{fig:Cor_I}
\end{figure}

Figure \ref{fig:Cor_I} plots $C(t)$ vs. $t$.
The value of $C(t)$ generally decreases from its initial value of 1 
with increasing $t$.
Because the sample number is restricted, 
$\Delta C(t)$ is large. 
We see that for difficult questions with $z_{0}(T)=50\% \pm 5\%$ and 
$60\% \pm 5\%$, $C(t)$ is positive for large values of $t$.
On the other hand, for easy questions with $z_{0}(T)=70\% \pm 5\%$ and 
$80\% \pm 5\%$, $C(t)$ decreases to zero with increasing $t$.
These results suggest that the system is in the two-peak phase for 
difficult questions. For $z_{0}(T)=70\% \pm 5\%$ and 
$80\% \pm 5\%$, an analysis of $q(z)$ showed that  
the system was in the one-peak phase \cite{Mor:2013}.

\begin{figure}[t]
\begin{center}
\begin{tabular}{c}
\includegraphics[width=8cm]{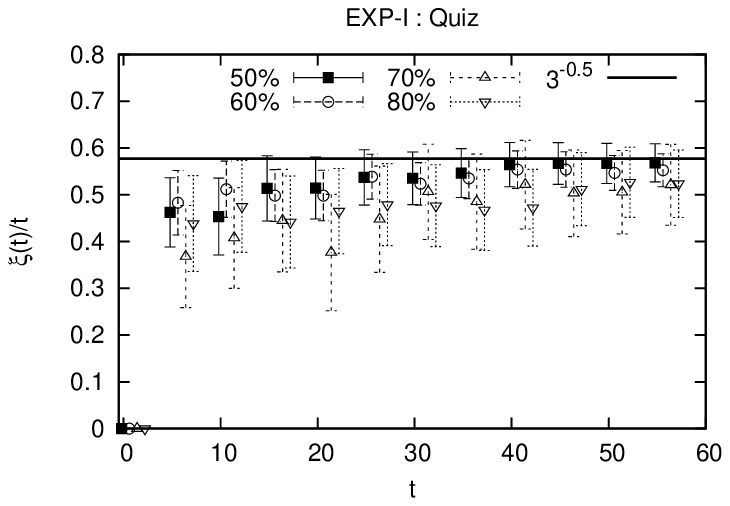} \\
\includegraphics[width=8cm]{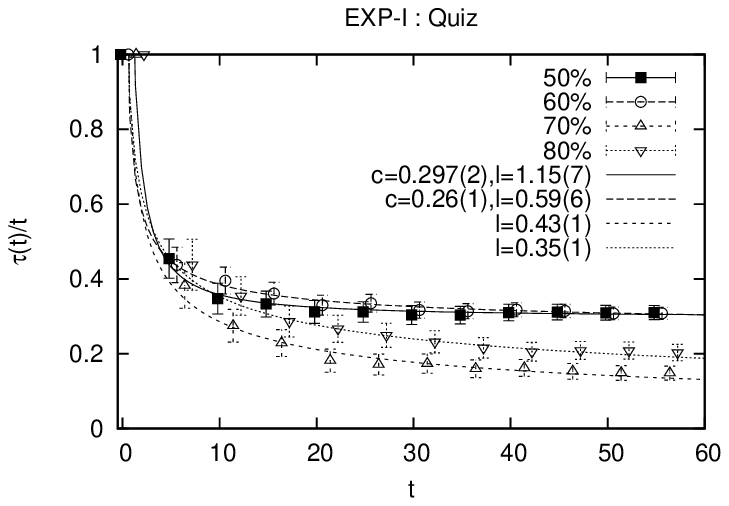} 
\end{tabular}
\end{center}
\caption{$\xi(t)/t$ and $\tau(t)/t$ vs. $t$ for EXP-I with the interval
$\Delta t=5$.
We also plot the fitted results for $\tau(t)/t$.
}
\label{fig:Xit_Taut_I}
\end{figure}

Figure \ref{fig:Xit_Taut_I} shows plots of $\xi(t)/t$ and $\tau(t)/t$ vs. $t$.
The standard errors for $\xi(t)/t$ are larger than those for $\tau(t)/t$ 
because $\xi(t)$ is calculated with 
the second moment $m_{2}(t)$. 
For large values of $t$, $\xi(t)/t$  takes $\sqrt{1/3}$ for difficult questions 
with $z_{0}(T)=50\% \pm 5\%$ and $60\% \pm 5\%$. The results suggest
 that the system is in the two-peak phase. 
For easy questions with $z_{0}(T)=70\% \pm 5\%$ and $80\% \pm 5\%$,
$\xi(t)/t \simeq 0.5$ for large values of $t$. 
As $\xi(t)/t\simeq \sqrt{l/l+2}$, $l \simeq 0.7$
 for easy questions. As $l$ is smaller than 1, 
the system is in the one-peak phase.

As the system is considered to be in the two-peak phase  
for $z_{0}(T)=50\% \pm 5\%$ and $60\% \pm 5\%$, 
we assume $\tau(t)/t=c+d\cdot t^{l-1}$ and estimate $c,l,d$ 
using the least square fit. 
We find that $c=0.297(2)$ for $z_{0}(T)=50\% \pm 5\%$ and 
$c=0.26(1)$ for $z_{0}(T)=60\% \pm 5\%$.
For $z_{0}(T)=70\% \pm 5\%$ and $80\% \pm 5\%$,
we assume $\tau(t)/t=d\cdot t^{l-1}$ and estimate $l$ and $d$.
We find that $l=0.43(1)$ for $z_{0}(T)=70\% \pm 5\%$ and $l=0.35(1)$ 
for $z_{0}(T)=80\% \pm 5\%$, which differ slightly 
from the value of $l\simeq 0.7$ estimated from $\xi(t)/t$.

\subsection{EXP-II: Urn choice case}

\begin{figure}[t]
\begin{center}
\includegraphics[width=8cm]{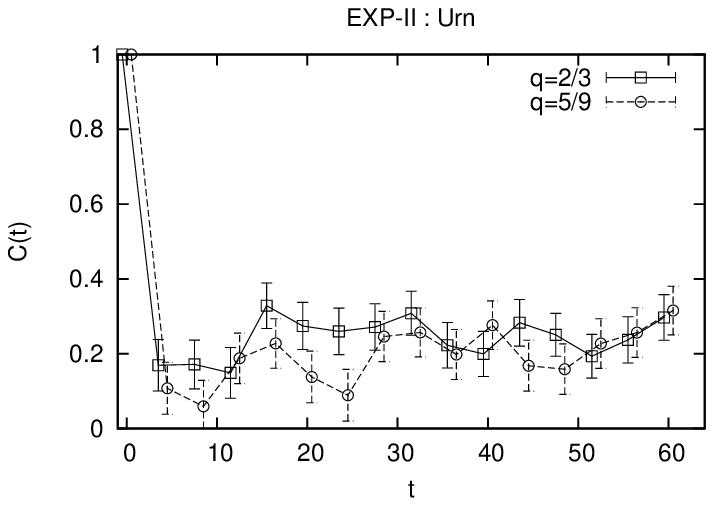} 
\begin{tabular}{cc}
\includegraphics[width=6cm]{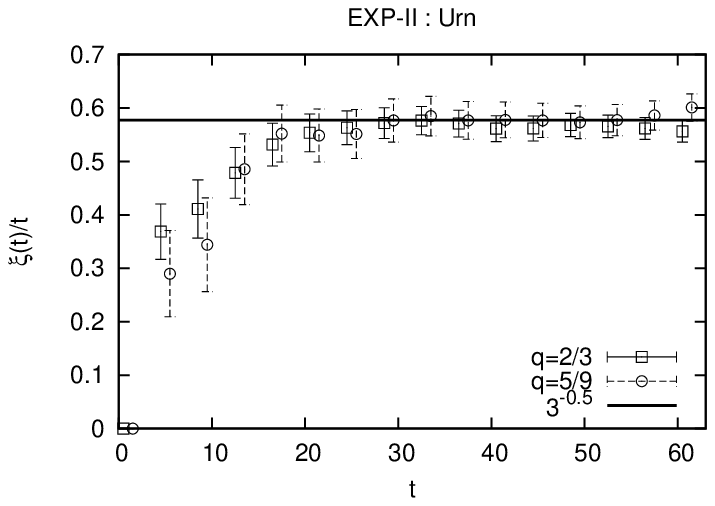} &
\includegraphics[width=6cm]{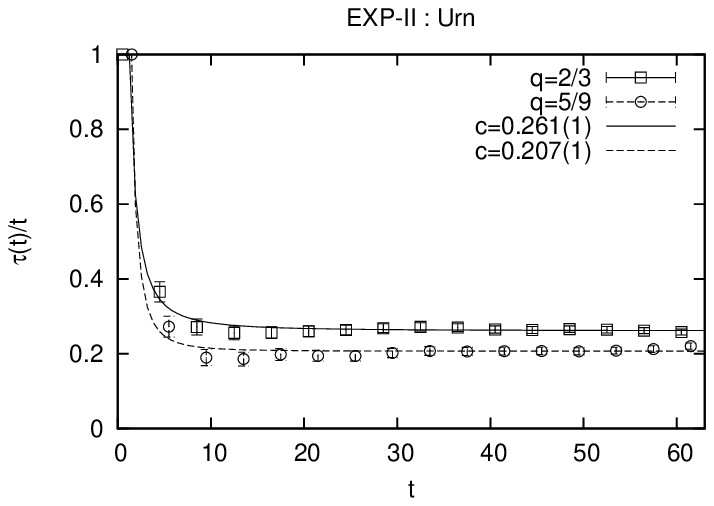} 
\end{tabular}
\end{center}
\caption{
$C(t)$, $\xi(t)/t$, and $\tau(t)/t$ vs. $t$ for EXP-II.
We use the symbol $\Box (\bigcirc)$ for $q=2/3 (5/9)$. 
We plot only data with the interval $\Delta t=4$. 
To see the behavior clearly, we slightly shift the data horizontally.}
\label{fig:Cor_Xit_Taut_II}
\end{figure}

Figure \ref{fig:Cor_Xit_Taut_II} shows plots of  
$C(t)$, $\xi(t)/t$, and $\tau(t)/t$
 vs. $t$ for $q\in \{2/3,5/9\}$.
As the number of samples is larger than that in EXP-I, 
 the standard errors are smaller than the symbols' 
 size for $\tau(t)/t$ and large $t$.
We see that $C(t)$ is positive for large values of $t$ for both cases of 
$q$, where $q\in \{2/3,5/9\}$.
In addition,  $\xi(t)/t$ for large values of $t$ converges to  
$\sqrt{1/3}$, and 
the exponent $l$ for $C(t)\sim t^{l-1}$ is almost one.
These results suggest that the system is in the two-peak phase for both 
values of $q$. 
We assume $\tau(t)/t=c+d\cdot t^{l-1}$ and estimate $c,l,d$ using the 
 least square fit. 
 We find that $c=0.261(1)$ for $q=2/3$ and $c=0.207(1)$ for $q=5/9$.

\section{Conclusion}
 We studied the self-correcting nature of information cascades.
 We proposed the use of the normalized correlation function $C(t)$,
 which shows how the first subject's choice is propagated to later 
 subjects and measures the strength of the domino effect in information cascades.
 $c\equiv \lim_{t\to\infty}C(t)>0$
 is a sufficient condition for a non-self-correcting  
 information cascade. 
 In this case, the domino effect continues infinitely.
 The system is in the two-peak phase, and the probability 
 that $z(t)$ converges to $z_{-}<1/2$ is positive.
 We used  data from two types of information cascade experiment:
 EXP-I, which used a general knowledge quiz, and 
 EXP-II, which used urns.
 The accuracy $q$ of the private signal is $q\in \{2/3,5/9\}$ in EXP-II. 
 We estimate $C(t)$ and its integrated quantities $\tau(t)$ and $\xi(t)$.
 In EXP-I, when the questions were difficult, $c>0$.
 In EXP-II, $c>0$ for both cases of $q$ where $q\in \{2/3,5/9\}$. In these cases,
 the system is non-self-correcting.
 
 We focus on the study of the non-self-correcting nature
 of information cascades.
 Although $c>0$ is a sufficient condition for a non-self-correcting cascade,
 $c=0$ is not a sufficient condition for 
 a self-correcting cascade. To verify this, one should study 
 the response function $q(z)$ and 
 count the number of solutions for $z=q(z)$. 
 Alternatively, it is necessary to study the 
 limit value of the variance of $z(t)$.
 If there is only one solution, $z_{+}>1/2$, or the limit value is zero,
 the system is self-correcting. In EXP-I, we studied 
 these points and  concluded that the system is self-correcting 
 for $z_{0}(T)=70\% \pm 5\%$ and $80\% \pm 5 \%$ \cite{Mor:2013}. 
 Our experiment for EXP-II and its analysis are under way \cite{Hin:2015}. 
 
\section*{Acknowledgment}
This work was supported by Grant-in-Aid for Challenging
Exploratory Research 25610109.


\begin{thebibliography}{4}
\bibitem{Cas:2009}Catellano, C., Fortunato, S., Loreto, V.: Statistical physics of social dynamics. Rev. Mod. Phys. 81, 591--646 (2009).

\bibitem{Gra:2014}Fern\'{a}ndez-Gracia, J., Sucheki, K., Ramasco, J.J., Miguel, M.S., Egu\'{i}luz, V.M.: Is the Voter Model a model for voters? Phys. Rev. Lett. 112, 158701--158705 (2014). 

\bibitem{Bik:1992}Bikhchandani, S., Hirshleifer, D., Welch, I.: A theory of fads, fashion, custom, and cultural changes as informational cascades. J. Polit. Econ. 100, 992--1026 (1992).

\bibitem{Dev:1996}Devenow, A., Welch, I.: Rational herding in financial economics. Euro. Econ. Rev. 40, 603--615 (1996).

\bibitem{Lee:1993}Lee, I.H.: On the convergence of informational cascades. J. Econ. Theory 61, 395--411 (1993).

\bibitem{And:1997}Anderson, L.R., Holt, C.A.: Information cascades in the laboratory. Am. Econ. Rev. 87, 847--862 (1997).

\bibitem{Kub:2004}K$\ddot{\mbox{u}}$bler, D., Weizs$\ddot{\mbox{a}}$cker, G.: Limited depth of reasoning and failure of cascade formation in 
the laboratory. Rev. Econ. Stud. 71, 425--441 (2004).

\bibitem{Goe:2007}Goeree, J.K., Palfrey, T.R., Rogers, B.W., McKelvey, R.D.: Self-correcting information cascades. Rev. Econ. Stud. 74, 733--762 (2007).

\bibitem{Mor:2012}Mori, S., Hisakado, M., Takahashi, T.: Phase transition to two-peaks phase in an information cascade voting experiment. Phys. Rev. E 86, 026109--026118 (2012). 

\bibitem{His:2011}Hisakado, M., Mori, S.: Digital herders and phase transition in a voting model. J. Phys. A 44, 275204--275220 (2011). 

\bibitem{His:2012}Hisakado, M., Mori, S.: Two kinds of phase transitions in a voting model. 
J. Phys. A, Math. Theor. 45, 345002--345016 (2012). 


\bibitem{Hil:1980}Hill, B., Lane, D., Sudderth, W.: A strong law for some generalized urn processes. Ann. Prob. 8, 214--226 (1980).

\bibitem{Hod:2004}Hod, S., Keshet, U.: Phase-transition in binary sequences with long-range correlations.
Phys. Rev. E 70, 015104--015109 (2004).

\bibitem{Mor:2015}Mori, S., Hisakado, M.: Finite-size scaling analysis of binary stochastic processes and universality classes of information cascade phase transition.  J. Phys. Soc. Jpn. 84, 054001--054013 (2015). 

\bibitem{Mor:2015a}Mori, S., Hisakado, M.: Correlation function for generalized Polya urns: Finite-size scaling analysis. arXi:arXiv:1501.00764.

\bibitem{Mor:2013}Mori, S., Hisakado, M., Takahashi, T.: Collective adoption of Max-Min strategy in an information cascade voting experiment. J. Phys. Soc. Jpn. 82, 084004--084013 (2013). 

\bibitem{Hin:2015}Hino, M., Hisakado, M., Takahashi, T., Mori, S.: Phase 
transition of generalized P\'{o}lya urn in Information cascade 
experiment. In preparation.
\end{thebibliography}
\end{document}